\documentclass{aa}  

\usepackage{graphicx}
\usepackage{txfonts}
\usepackage{natbib}
\usepackage{xcolor}
\newcommand{\bdot}{ \boldsymbol{\cdot}}
\newcommand{\bnabla}{ \boldsymbol{\nabla} }
\newcommand{\xhat}{ \boldsymbol{\hat{x}} }
\newcommand{\yhat}{ \boldsymbol{\hat{y}} }
\newcommand{\rsun}{R}

\newcommand{\ii}{ {\rm i} }
\newcommand{\id}{ {\rm d} }

\newcommand{\bu}{ \boldsymbol{u} }

\newcommand{\LG}[1]{\textcolor{black}{#1}}

\begin{document} 
\title{Effect of latitudinal differential rotation on solar Rossby waves:  Critical layers,  eigenfunctions, and momentum fluxes\\ in the equatorial $\beta$ plane }

\titlerunning{Effect of latitudinal differential rotation on solar Rossby waves}

\author{
L. Gizon\inst{1,2,3}
\and D. Fournier\inst{1}
\and M. Albekioni\inst{2,4,5}
       }
\institute{Max-Planck-Institut f\"ur Sonnensystemforschung, Justus-von-Liebig-Weg 3, 37077 G\"ottingen, Germany\\ \email{gizon@mps.mpg.de}
             \and
             Institut f\"ur Astrophysik, Georg-August-Universit\"at G\"ottingen,  Friedrich-Hund-Platz 1, 37077 G\"ottingen, Germany
             \and
             Center for Space Science, NYUAD Institute, New York University Abu Dhabi, Abu Dhabi, UAE
             \and
             Ilia State University, Kakutsa Cholokashvili Ave. 3/5, Tbilisi 0162, Georgia  
             \and
             Evgeni Kharadze Georgian National Astrophysical Observatory, Abastumani, Adigeni 0301, Georgia
               }
   \date{Submitted on 29 May 2020 to Astron. Astrophys.}

 
  \abstract
   {Retrograde-propagating waves of vertical vorticity with longitudinal wavenumbers between 3 and 15 have been observed on the Sun with a dispersion relation close to that of classical sectoral Rossby waves.
   The observed vorticity eigenfunctions  are symmetric in latitude, peak at the equator, switch sign near $20^\circ$\,--\,$30^\circ$, and decrease at higher latitudes.}
   {We search for an explanation that takes into account solar latitudinal differential rotation.} 
   {In the equatorial $\beta$ plane, we study the propagation of linear Rossby waves (phase speed $c <0$) in a parabolic zonal shear flow, $U  = - \overline{U}\ \xi^2<0$,  where $\overline{U} = 244$~m/s and  $\xi$ is the sine of latitude. 
   }
   {In the inviscid case, the eigenvalue spectrum is real and  continuous and the velocity stream functions are singular at the critical latitudes where $U = c$. 
   We add eddy viscosity in the problem to account for wave attenuation.
   In the viscous case, the stream functions are solution of a fourth-order modified Orr-Sommerfeld equation. Eigenvalues are complex and discrete. 
   \LG{For reasonable values of the eddy viscosity corresponding to supergranular scales and above (Reynolds number $100 \le  Re \le 700$), all modes are stable.}
   At fixed longitudinal wavenumber, the least damped mode is a symmetric mode with a real frequency close to that of the classical Rossby mode, which we call the R mode.  For $Re \approx 300$, the  attenuation and the real part of the eigenfunction is in qualitative agreement with the observations (unlike the imaginary part of the eigenfunction, which has a larger amplitude in the model).  
  }
   {Each longitudinal wavenumber is associated with a  latitudinally symmetric R mode trapped  at low latitudes by solar differential rotation.  In the viscous model, R modes transport  significant angular  momentum  from the dissipation layers towards the equator.
   }


\maketitle
  
\newpage


\section{Introduction}

In the earth atmosphere, \citet{Rossby1939} waves are global-scale  waves of radial vorticity that propagate in the direction opposite to rotation (retrograde).  They find their origin in the conservation of  vertical absolute vorticity, i.e. the sum of planetary and wave vorticity \LG{\citep[see, e.g.,][]{Platzman1968, Gill1982}}.

Equatorial Rossby waves have recently been observed on the Sun  with longitudinal wavenumbers in the range $3\le m \le 15$ \citep{LOE18,LIA19}. In the corotating frame, 
their dispersion relation is close to that of classical sectoral ($l=m$) Rossby waves,  $\omega = - 2 \Omega/(m+1)$, where $\Omega/2\pi=453.1$~nHz is the equatorial rotation rate.

The observed  variation of the eigenfunctions with latitude, however, differs noticeably  from $P_m^m(\sin\lambda) \propto (\cos\lambda)^m$ where $\lambda$ is latitude, which is the expected answer for  sectoral  modes in a uniformly rotating sphere \citep[e.g.,][]{Saio1982, Damiani2020}. Instead,  the observed eigenfunctions have real parts that peak at the equator, switch sign near $20^\circ$\,--\,$30^\circ$, and decay at higher latitudes \citep{LOE18}. Their imaginary parts are small \citep{PRO20}.

An ingredient that is obviously missing in models of solar Rossby waves is latitudinal differential rotation. The Sun's rotation rate decreases fast with latitude: the difference between the rotation rate at mid latitudes and at the equator is not small compared to the frequencies of the Rossby waves of interest. For $m$ larger than 5, we will show that there exists a critical latitude at which the (negative) wave frequency equals the (negative) differential rotation rate counted from the equator.

To capture the essential physics while keeping the problem simple, we choose to work in two dimensions and in the equatorial $\beta$ plane. This simplification is acceptable for wavenumbers that are large enough (say $\ge 6$).

The stability and dynamics of parabolic (Poiseuille) shear flows in the presence of critical layers was summarized by, e.g.,  \citet{Drazin}. \citet{KUO49} included the $\beta$ effect in the problem. In the inviscid case, critical layers lead to a singular eigenvalue problem \citep[see, e.g.,][]{BAL95}. The stream functions are continuous but not differentiable (Sect.~\ref{sect:inviscidNum}), thus they cannot be compared directly to actual observations of the vorticity. Since we also wish to explain the lifetime of the modes \citep{LOE18}, we introduce a viscous term in the Navier-Stokes equations  to model damping by turbulent convection. As shown in Sect.~\ref{sec:OSeq}, this leads to a new equation for the stream function: an Orr-Sommerfeld equation with coefficients modified by the $\beta$ effect. The viscosity removes singularities and the eigenfunctions are regular across the viscous critical layer (see Sect.~\ref{sec:layers}). To solve the eigenvalue problem accurately, we use a numerical method based on the Chebyshev decomposition proposed by \citet{ORS71}. As shown in Sect.~\ref{sec:viscous}, the eigenvalue spectrum includes a symmetric Rossby mode in addition to the other  modes that are known to exist in the $\beta=0$ case  \citep{Mack1976}. Note that in the present paper we focus on the eigenvalue problem. We do not discuss the nonlinear dynamics in the critical layers, which would require solving the nonlinear evolution equation \citep[e.g.,][]{Stewartson1977}.

In addition to the practical advantages of studying 2D Rossby waves in the $\beta$ plane,  the physics of this problem has been extensively discussed in earth and planetary sciences. In the earth atmosphere and oceans,  Rossby waves encounter critical layers \citep[see][]{Frederiksen1988,VAL06,Boyd2018}. They play a role in the global dynamics by transporting angular momentum via Reynolds stresses and they modify the mean flow \citep[e.g.,][]{Bennett1971,Webster1973,Geisler1974}.

Our model can be further extended to include the  effect of the solar meridional flow using \citet{ORS71}'s method, see Sect.~\ref{sec:merid}.
In Sect.~\ref{sec:obs} we compare the predictions of the model to solar observations of the mode frequencies and  damping rates and to observations of the vorticity eigenfunctions.
Finally, we discuss in Sect.~\ref{sec:reynolds} implications of the model for  angular momentum transport  and the dynamics of solar differential rotation.



\section{Waves in a sheared zonal  flow: Equations of motion in the equatorial $\beta$ plane}

\label{sec:OSeq}

In the equatorial $\beta$ plane, we study the propagation of two-dimensional Rossby waves through  a steady  zonal flow representative of solar differential rotation. 
 Several $\beta$-plane approximations have been proposed \citep[see, e.g.,][]{Dellar2011}.
Here we choose the  sine transformation:
   \begin{align}
    x &= \rsun \phi , 
    \\
    y &= \rsun \sin \lambda, \qquad -\rsun \le y \le \rsun,
\end{align}
where $\phi$ is longitude, $\lambda$ is latitude, and $\rsun=696$~Mm is the solar radius. 
The $x$ coordinate increases in the prograde direction and the $y$ coordinate increases northward.
To first order in $y/R$, the  $x$ and $y$ components  of the velocity vector  in the $\beta$ plane are respectively equal to their $\phi$ and $\lambda$ components  on the sphere \citep{Ripa1997}.

\begin{figure}[t]
   \includegraphics[width=\linewidth]{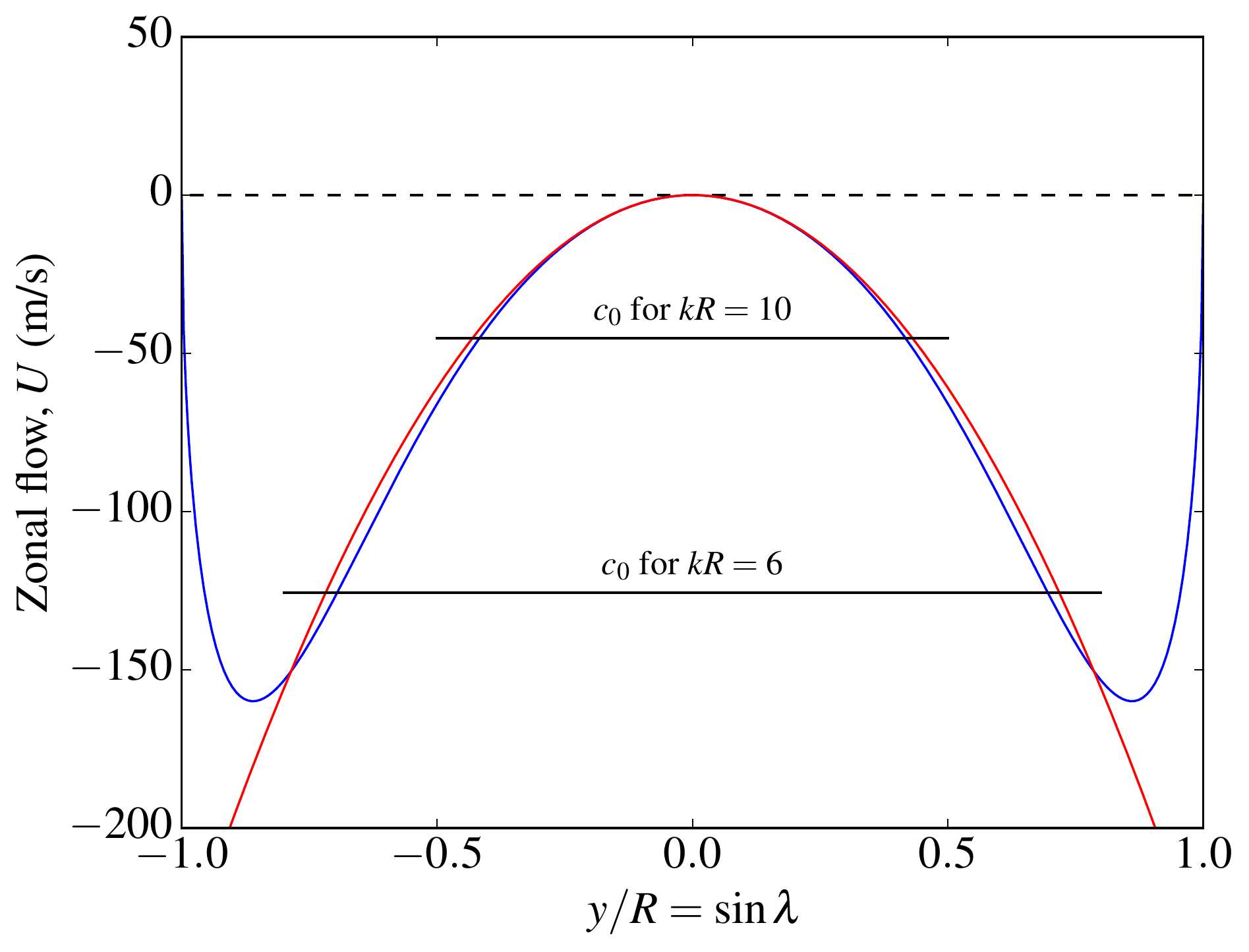} 
\caption{Parabolic zonal flow $U$ (red curve, Eq.~\ref{eq:flow}) in the frame rotating at the equatorial rotation rate, which approximates the Sun's rotational velocity at the photosphere (blue curve).
The horizontal black lines indicate the phase speed of Rossby waves, $c_0=-(\beta-U'')/k^2$,  for longitudinal wavenumbers $kR=6$ and $kR=10$. There are  critical latitudes  where $U =c_0$.
}
\label{fig:flow}
\end{figure}

The total velocity is the sum of the background zonal flow $U(y) \xhat$
  and the horizontal wave velocity $\bu (x,y,t)$ with
\begin{equation}
  \bu (x,y,t) = u_x(x,y,t) \xhat + u_y(x,y,t) \yhat. 
\end{equation}
By choice, $U(0)=0$ at the equator. The latitudinal shear is specified by the parabolic (Poiseuille) profile
\begin{equation}
 U (y) =  - \overline{U}\ (y/\rsun)^2 ,  \label{eq:flow}
\end{equation}
where the amplitude  $\overline{U}$ is chosen to approximate \LG{the Sun's surface differential rotation at low and mid latitudes. 
From the `standard' solar angular velocity profile given by \citet{Beck2000}, 
$\Delta\Omega = - 0.35 [ (y/R)^2 + (y/R)^4]\ \mu$rad/s, we find that the value $\overline{U} = 244$~m/s is a good choice, see Fig.~\ref{fig:flow}.}

The two-dimensional Navier-Stokes equations in the equatorial  $\beta$ plane are
\begin{align}
& \left(\frac{\partial}{\partial t}  + U \frac{\partial}{\partial x} \right)  u_x + u_y U' +  \bu\bdot\bnabla  u_x   = - \frac{1}{\rho} \frac{\partial p}{\partial x} + \nu \Delta u_x
+ f u_y
,
\\
& \left( \frac{\partial}{\partial t}  +  U \frac{\partial}{\partial x} \right)  u_y +  \bu\bdot\bnabla  u_y    = - \frac{1}{\rho}\frac{\partial p}{\partial y}  + \nu  \Delta u_y  - f u_x ,
\end{align}
where $\nu$ is the viscosity and the equatorial Coriolis parameter is $f=\beta y$ with $\beta =2\Omega/\rsun$. 
The prime denotes a derivative, for example $U' = {\rm d} U/ {\rm d} y$.

To enforce mass conservation, we introduce the stream function $\Psi$ such that
\begin{equation}
u_x=\frac{\partial \Psi}{\partial y}  , \qquad  u_y =-\frac{\partial \Psi}{\partial x}.
\end{equation}
Assuming a barotropic fluid, we can  eliminate the pressure term by combining the two components of the momentum equation, to obtain
\begin{equation}
\left(  \frac{\partial}{\partial t}    +  U \frac{\partial}{\partial x}\right) \Delta \Psi
  +  \left( \beta - U'' \right) \frac{\partial\Psi}{\partial x}
  +  (\bu\bdot\bnabla  ) \Delta \Psi
=     \nu\ \Delta^2  \Psi  . \label{eq:initialTime}
\end{equation}

\section{Critical latitudes}
\label{sec:layers}

\subsection{Linear inviscid case: critical points}

In the linear inviscid case, we search for wave solutions of the form
\begin{equation}
    \Psi (x,y,t) =  {\rm Re} \left\{ \psi(y) \exp[{\ii(k x - \omega t )}] \right\}.
\end{equation}
Dropping the nonlinear and  viscous terms, Eq.~\eqref{eq:initialTime}
becomes the Rayleigh-Kuo equation \citep{KUO49}:
\begin{equation}
\left( U-c  \right) (-k^2\psi+ \psi'')
  +   \left( \beta - U'' \right) \psi 
=   0, \label{eq:RayleighKuo}
\end{equation}
where $c=\omega/k$ is the phase speed.
The above equation differs from Rayleigh's (\citeyear{RAY79}) equation only through the $\beta$ term.
It can be  rewritten  as a Helmholtz equation:
\begin{equation}
 \psi'' + K(y)  \psi =   0, 
 \label{eq:helm}
\end{equation}
with
\begin{equation}
K(y) = -k^2 +  \frac{\beta-U''}{U-c}.
\label{eq:q2}
\end{equation}
The squared latitudinal wavenumber, $K(y)$, is singular at the critical points $y=\pm y_c$ such that $U(y_c) = c$.
The critical points divide the low-latitude region where the solution is locally oscillatory ($K>0$ for $|y|<y_c$) from the high-latitude regions where it is locally evanescent ($K<0$ for $|y|>y_c$).

Equation~\eqref{eq:RayleighKuo}, supplemented by boundary conditions $\psi(\pm \rsun)=0$,  is an eigenvalue problem that can be solved in the complex plane. It admits a continuum of neutral modes with real eigenfrequencies. According to Rayleigh's theorem \citep[adapted for the Rayleigh-Kuo equation, see][]{KUO49}, there is no discrete mode because $\beta-U'' \neq 0$ everywhere. We can thus solve the initial value problem given by Eq.~\eqref{eq:RayleighKuo} for any particular real value  of $c$  to obtain the associated eigenfunction \LG{\citep[e.g.,][]{DRA66,Drazin1982, BAL95}}. These eigenfunctions are singular at the critical points.

Because $U''$ is constant in our problem and $U(0)=0$, Eq.~\eqref{eq:q2} implies that each mode can be associated with a value of  $K(0)$ such that
\begin{equation}
    c =  - \frac{\beta-U''}{k^2+K(0)}  .
\end{equation}
For equatorial propagation, $K(0) = 0$ is a natural choice,
and we may consider the eigenvalue  
\begin{equation}
c_0 = - (\beta-U'')/k^2
\label{eq.c0}
\end{equation}
as an example.  
In our case, $\beta-U'' = 1.12 \beta$, so waves propagate faster than   in the no-flow case.
The critical points $y = \pm y_c$ where $U(\pm y)=c_0$ are given by
\begin{equation}
    k y_c   
    =  \sqrt{\beta R^2/\overline{U} + 2}
    = 4.31 .
\end{equation}
Thus, for $kR>4.31$, there are  critical latitudes  at $\lambda =\pm \lambda_c$, such that 
\begin{equation}
    \lambda_c = \arcsin (4.31 / kR ) .
\end{equation}
To obtain the eigenfunctions, Eq.~(\ref{eq:RayleighKuo}) should be solved separately for $|y| < y_c$ and $|y| > y_c$. The analytical and numerical solutions are discussed in Sect.~\ref{sect:inviscidNum}. The stream function is continuous  ($u_y = -\ii k \psi$ is continuous), however its first and second derivatives diverge at the critical layer \citep[see, e.g.,][]{Haynes2003}.

\subsection{Viscous critical layer}

Bulk viscosity removes singularities. The  linear viscous equation  for $\psi$ is 
\begin{equation}
\left( c-U  \right) (-k^2\psi +\psi'')
  -  \left( \beta - U'' \right) \psi 
=   \frac{ \ii\nu}{ k} (k^4\psi -2 k^2 \psi''+ \psi''''). \label{eq:OrrSommerfeld}
\end{equation}
For $\beta=0$ we recognize the Orr-Sommerfeld equation \citep{ORR07,SOM09}. Equation~(\ref{eq:OrrSommerfeld}) is a fourth-order differential equation, which requires four boundary conditions, e.g.  $\psi(\pm \rsun) = 0$ and $\psi'(\pm \rsun) = 0$ for a no-slip boundary condition. 
The  critical layer of the inviscid case is replaced by a viscous critical layer  around $y=\pm y_c$.  
The width of this viscous layer, $\delta$, is obtained by balancing the dominant viscous term with the dominant  term on the left-hand side, $(U-c)\psi'' \sim \nu \psi''''/ k $. Close to the viscous layer, we write $\id/\id y \sim 1/\delta$ and  $U-c \approx U'(y_c) (y-y_c) \sim (\overline{U} y_c/R^2) \delta$, so that the width of the viscous layer  is approximately
\begin{equation}
 \delta/\rsun    \sim (k y_c\ Re)^{-1/3},
\end{equation}
where 
\begin{equation}
    Re = \rsun \overline{U} / \nu
\end{equation} is the Reynolds number.


The width of the layer is controlled by $Re$.  As discussed by \cite{Ruediger1989}, $\nu$ should be understood as an eddy viscosity due to turbulent convection. As an example, we may estimate the Reynolds number associated with solar supergranulation. For supergranulation, the  turbulent viscosity  $\nu \approx 250$~km$^2$/s \citep{SIM97, DUV00}  implies $Re \approx 700$
and $\delta \approx 0.07 R$ for $k \rsun = 10$. Not surprisingly, the width of the viscous layer  is comparable in this case to the spatial scale of supergranulation.

\subsection{Nonlinear critical layer}
   
In order to assess whether it is legitimate to drop the nonlinear term $(\bu\bdot\bnabla  ) \Delta \Psi$ in Eq.~\eqref{eq:initialTime}, we estimate
the width of the nonlinear critical layer  $\delta_{NL}$. It is obtained by balancing the advection term $k(U-c)\psi''$ and the nonlinear term $u_y  \psi'''$.
We find
\begin{equation}
\delta_{NL} / \rsun \sim \left(u_{\rm max}/\overline{U}\right)^{1/2} (k y_c)^{-1/2} ,
\end{equation}
where $u_{\rm max}$ is a characteristic velocity amplitude for the Rossby waves. 
On the Sun, \citet{LIA19} measured  $u_{\rm max} \approx 2$~m/s
for the maximum latitudinal velocity of a mode at the equator.
For $k \rsun = 10$, the width of the nonlinear critical layer is  $\delta_{NL} \approx 0.04 R$.

Introducing the threshold $Re_* = (u_{\rm max}/\overline{U})^{-3/2} (k y_c)^{1/2}$, the ratio between the widths of the viscous and nonlinear critical layers is
\begin{equation}
    \delta / \delta_{NL} \approx  (Re/Re_*)^{-1/3} . 
\end{equation}
For $Re < Re_*$  the critical layer is linear and dominated by dissipation over the width  $\delta$. 
For $kR=10$, we find $Re_* \approx 3000$, which is much larger than the Reynolds number $Re \approx 700$ estimated in the previous section for solar supergranulation. Hence,  it is not unreasonable to study the linear problem --- as we do in the remainder of this paper. However, we caution that there is some uncertainty about the appropriate value for the  viscosity.

\section{Inviscid modes of oscillation} \label{sect:inviscidNum}

In the inviscid case, the spectrum for the Rayleigh-Kuo equation, Eq.~\eqref{eq:RayleighKuo}, is real and continuous. We fix the value of $c$ to $c_0$ (Eq. \ref{eq.c0}) and compute the corresponding eigenfunctions. There are two singular solutions, both real: a solution that is symmmetric in latitude and an antisymmetric solution. These solutions can be obtained by solving the equation analytically or numerically in two distinct intervals, $0\le y< y_c$ and $y_c \le y \le R $.

In the inner region, we solve the ODE with the conditions $\psi(0)=1$ and $\psi'(0)=0$ to obtain the symmetric solution. 
For the region $y>y_c$, we impose continuity with the inner solution and use the boundary condition $\psi(R)=0$. 
The symmetric solution can be expressed as a series in each regions. We write  $\psi(y) =  \sum_{p\ge 0} a_p \xi^{2p}$ with $\xi=y/R$ for $|y|<y_c$ and $\psi(y) = \sum_{p\ge 1} b_p (\xi^2-1)^p$ for $|y|>y_c$.
The coefficients $a_p$ and $b_p$  are computed by recurrence. Setting
 $\kappa = kR$ and $\xi_c = y_c/R$, the inner solution is given by
 \begin{align}
& a_0 =  1  , \\
& a_1 = 0  , \\
& a_p = \frac{2(p-1)(2p-3)a_{p-1} - \kappa^2 a_{p-2}}{2 \xi_c^2 p (2p-1) } \; \textrm{  for } p\ge 2 .
\end{align}
For the outer solution,
\begin{align}
 b_2 = &- b_1/4  , \\
 b_3 =& - \frac{[\kappa^2 + 3(1-\xi_c^2)]b_2}{6(1-\xi_c^2)} ,    \\
 b_p =& - \frac{ [(1-\xi_c^2)(2p-3)+2(p-2)]b_{p-1} }{2(1- \xi_c^2)  p } \nonumber \\
& + \frac{[\kappa^2 - 2(p-2)(2p-5)] b_{p-2}+\kappa^2b_{p-3}}{4(1- \xi_c^2) (p-1) p } \; \textrm{  for } p\ge 4 .
\end{align}
The coefficient $b_1$ is chosen such that the solution is continuous at the critical point. 

To validate the series solution, we also solved the problem numerically. In the inner region, we have  an initial value problem that can be solved using classical libraries, for example  \verb+odeint+ from \verb+SciPy+. In the outer region, the problem is a boundary value problem that can be converted into an initial value problem using the shooting method \citep{KEL68}.

\begin{figure}
       \centering
       \includegraphics[width=\linewidth]{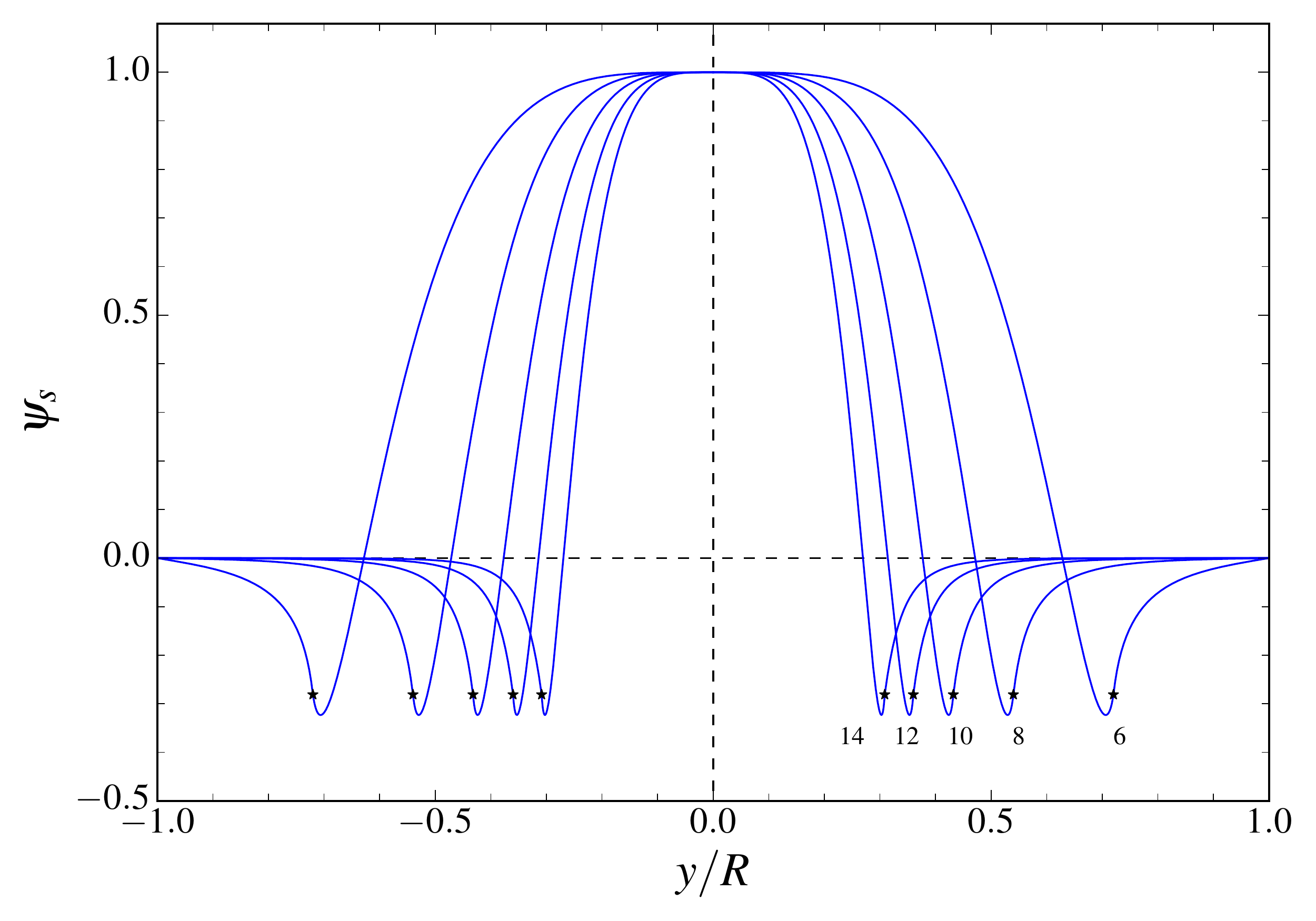}
       \caption{Symmetric inviscid eigenfunctions for  the eigenvalue $c_0=-(\beta-U'')/k^2$ and $k R =6, 8, 10, 12$ and $14$.  The stars mark the critical points.}
       \label{fig:inviscidS}
   \end{figure}

We find an excellent agreement between the analytical and the numerical solutions, thus we only plot the analytical solution in Fig.~\ref{fig:inviscidS} for $k R = 10$. 
The symmetric eigenfunction  $\psi_s$  switches sign before the critical latitude and evanesces  above it. The location of the critical layer is close to the zero-crossing of the observed vorticity eigenfunctions from \citet{LOE18} and \citet{PRO20}.
Unfortunately, in the inviscid case, the vorticity $\zeta =  k^2 \psi- \psi''$ diverges at the critical latitude and thus the  comparison with the  observations is difficult.

\begin{figure}
       \centering
       \includegraphics[width=\linewidth]{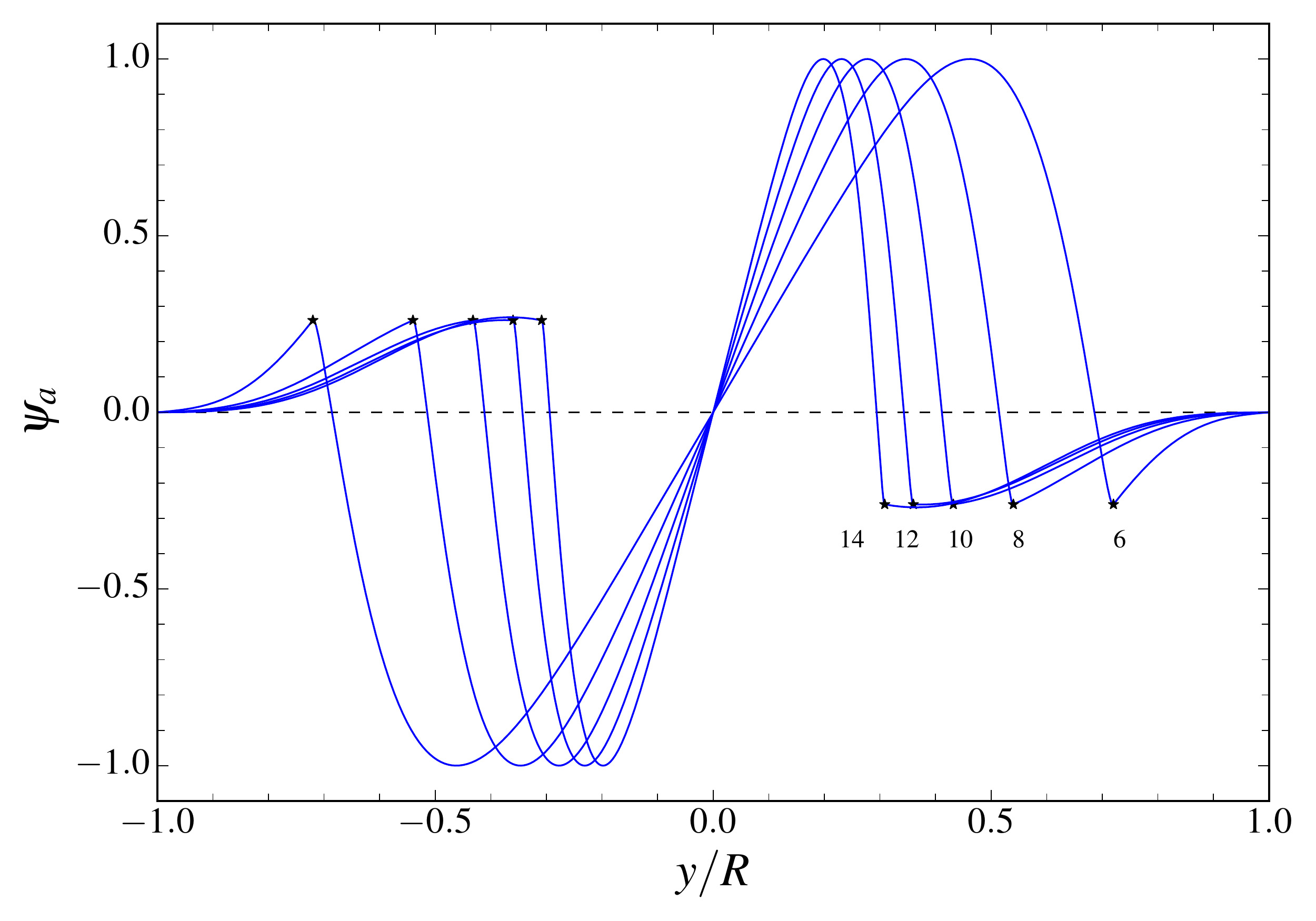}
       \caption{Antisymmetric inviscid eigenfunctions for $c=c_0$ and  $k R =6, 8, 10, 12$ and $14$.  The stars mark the critical points.}
       \label{fig:inviscidA}
   \end{figure}

 \begin{figure*}
       \centering
       \includegraphics[width=0.8\linewidth]{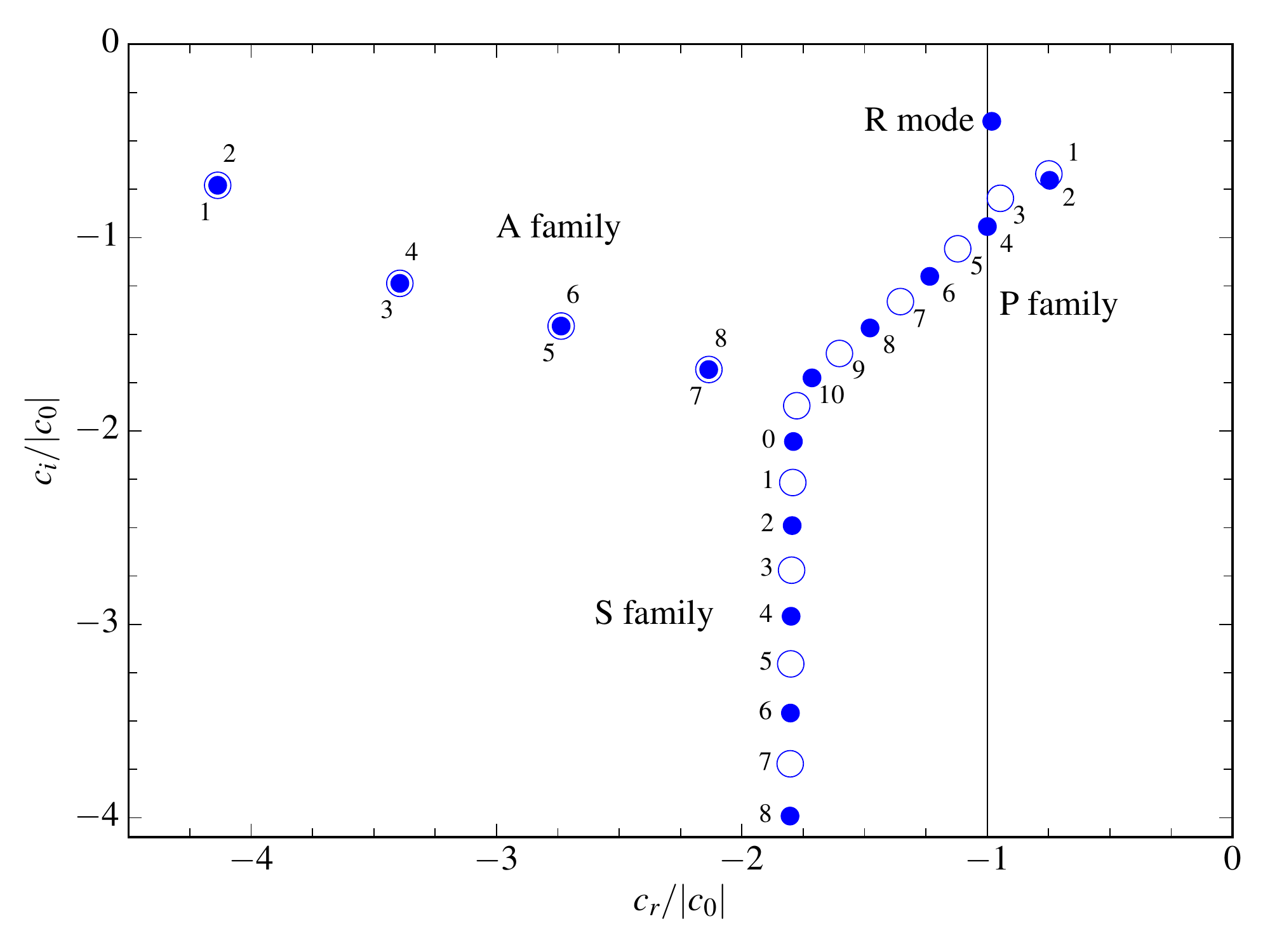}
       \caption{Eigenvalues $c=c_r + \ii c_i$ in the complex plane for $k R =10$ and $Re = 300$, normalized by  the reference eigenvalue $|c_0|=(\beta-U'')/k^2$.
       The R mode is clearly identifiable (the least damped mode), as well as the three families of modes: the wall modes (A family), the center modes (P family), and the damped modes (S family). The modes with symmetric eigenfunctions are shown with full circles and the ones with antisymmetric eigenfunctions with open circles. 
      All modes have $c_r<0$ and $c_i<0$, i.e. they are stable and propagate in the retrograde direction. The vertical line corresponds to the phase speed of the standard Rossby wave, $c_r=c_0$. Modes in each family are labelled with integers, which  increase with attenuation.}
       \label{fig:eigenvalues}
   \end{figure*}

 \begin{figure*}
       \centering
       \includegraphics[width=\linewidth]{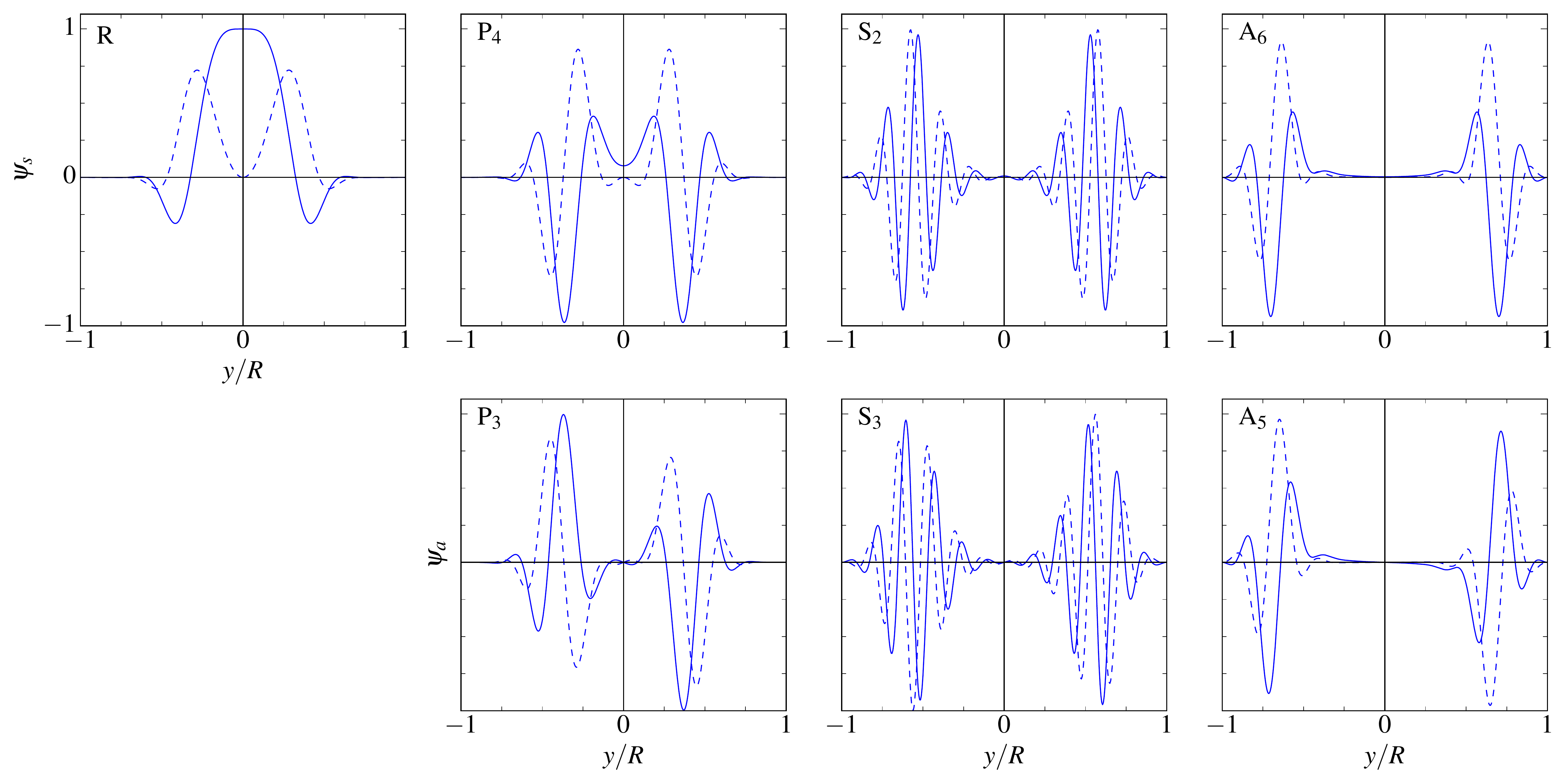}
       \caption{Eigenfunctions for the symmetric modes R, P$_4$, S$_2$, and A$_6$ (top row, $\psi_s$)
       and for the antisymmetric modes P$_3$, S$_3$, and A$_5$ (bottom row, $\psi_a$). See Fig.~\ref{fig:eigenvalues} for the position of the corresponding eigenvalues in the complex plane. The real and imaginary parts are plotted with solid and dashed lines respectively. The modulus of the eigenfunctions is normalized to one. By choice, all imaginary parts are  zero at  the equator.}
       \label{fig:eigenfunctions}
   \end{figure*}

\begin{figure*}
       \centering
       \includegraphics[width=\linewidth]{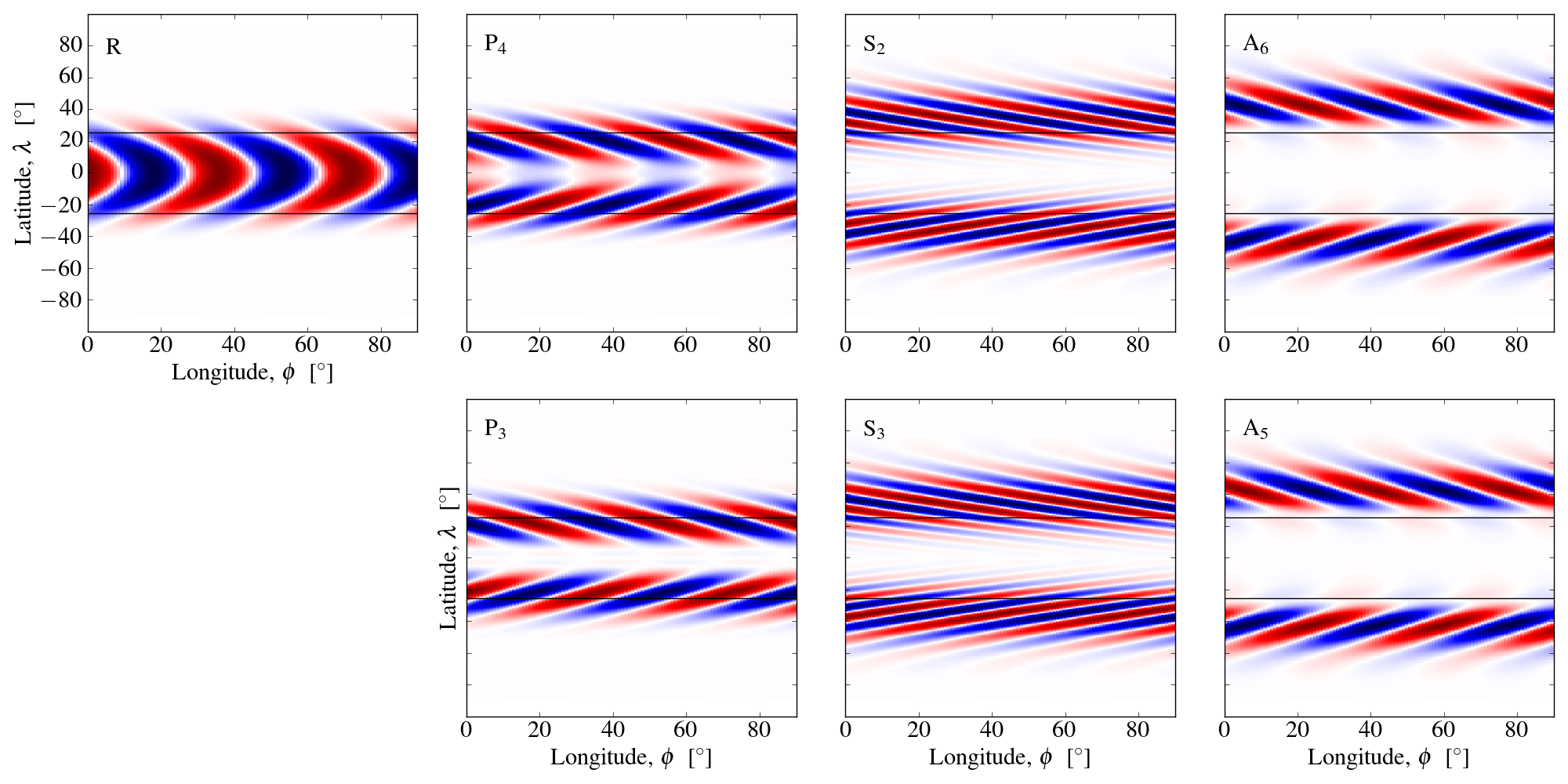}
       \caption{Stream functions in real space for all the modes shown in Fig.~\ref{fig:eigenfunctions}. Horizontal black lines show the central latitudes of the viscous layers, $\lambda = \pm 25^\circ$ for $kR=10$. The R mode (top left panel) is confined to the equatorial region between the viscous layers.}
       \label{fig:chevrons}
   \end{figure*}

We note that for each eigenvalue $c=c_0$ there exists also an antisymmetric eigenfunction, $\psi_a$. This solution can be obtained like above, but  with different boundary conditions at the equator. We set $\psi_a(0)=0$ and $\psi'_a(0)=\psi'_0$, where $\psi'_0$ can be chosen to control the maximum value of $\psi_a$, for example unity. The other conditions are the same as before, i.e. $\psi_a(\pm R)=0$ and $\psi_a$ continuous at the critical point.
The antisymmetric solution can be expanded as  $\psi_a(y) =  \sum_{p\ge 0} A_p \xi^{2p+1}$  for $|y|<y_c$ and $\psi_a(y) = \sum_{p\ge 1} B_p \xi (\xi^2-1)^p$ for $|y|>y_c$. 
The coefficients $A_p$ an $B_p$ are again obtained by recurrence. Figure~\ref{fig:inviscidA} shows example antisymmetric eigenfunctions. 

\section{Viscous modes of oscillation}
\label{sec:viscous}

\subsection{Numerical method}
\label{sec.numericalsolution}

In order to remove the singularities at the critical latitudes, we now include viscosity. The viscosity is specified through the Reynolds number, $Re$, which is a free parameter in our problem. For example,  $Re = 700$ for supergranular turbulent viscosity. 

To facilitate the numerical resolution of the modified  Orr-Sommerfeld equation,  Eq.~\eqref{eq:OrrSommerfeld}, it is convenient to introduce dimensionless quantities. We define  $\xi = y / \rsun$,  $\kappa = k \rsun$, and $\hat{\beta} = \beta \rsun^2 / \overline{U}$. The dimensionless eigenvalues $\hat{c} = c / \overline{U}$ and eigenfunctions $\hat{\psi}(\xi) = {\psi(y)}/({\rsun \overline{U}})$ solve the equation
\begin{equation} 
 (\hat{c} + \xi^2) \hat{D}\hat{\psi} -  (\hat{\beta} +2) \hat{\psi}  =  \ii (\kappa \ Re)^{-1}\
    \hat{D}^2 \hat{\psi},
\label{eq:orr}
\end{equation}
where $\hat{D} = - \kappa^2 + \id^2/\id \xi^2$ is the Laplacian and the prime now denotes derivation with respect to $\xi$.
For $\overline{U} =244$~m/s,  we have  $\hat{\beta}  = 16.4$.  We consider values of the dimensionless longitudinal wavenumber  in the range    $8 \leq \kappa  \leq  15$. 

We follow the numerical approach of \citet{ORS71}, originally developed for the Orr-Sommerfeld eigenvalue problem. We use the Matlab package {\verb+Chebfun+} to project functions onto Chebyshev polynomials and to compute spatial derivatives analytically \citep{DRI14}. This package also provides practical tools to solve differential equations and eigenvalue problems (only a few lines of codes are needed).

To obtain the symmetric solutions, we solve the above eigenvalue  problem on $[0,1]$ with the boundary conditions
\begin{equation}
\hat{\psi}'_s(0) =  \hat{\psi}_s'''(0) = 0 
\quad \text{and} \quad 
     \hat{\psi}_s(1) =  \hat{\psi}'_s(1) = 0 .
\end{equation}
The antisymmetric solutions are obtained by setting
\begin{equation}
\hat{\psi}_a(0) =  \hat{\psi}_a''(0) = 0 
\quad \text{and} \quad 
     \hat{\psi}_a(1) =  \hat{\psi}'_a (1) = 0 .
\end{equation}
In both cases, the numerical solutions (eigenvalues and eigenfunctions)  are complex. 

 \begin{figure*}[t]
       \centering
       \includegraphics[width=0.48\linewidth]{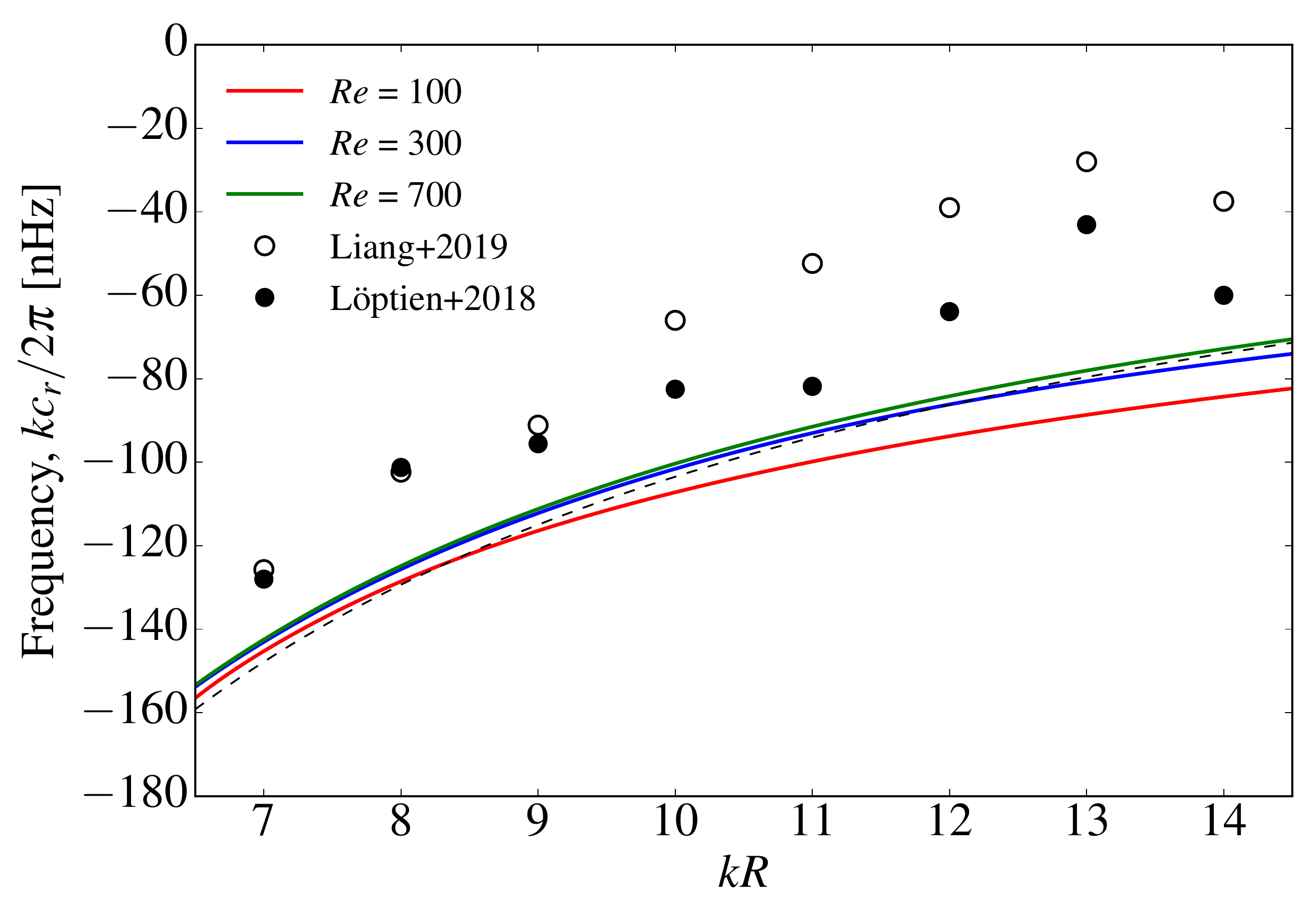} 
       \hspace*{0.2cm}
       \includegraphics[width=0.48\linewidth]{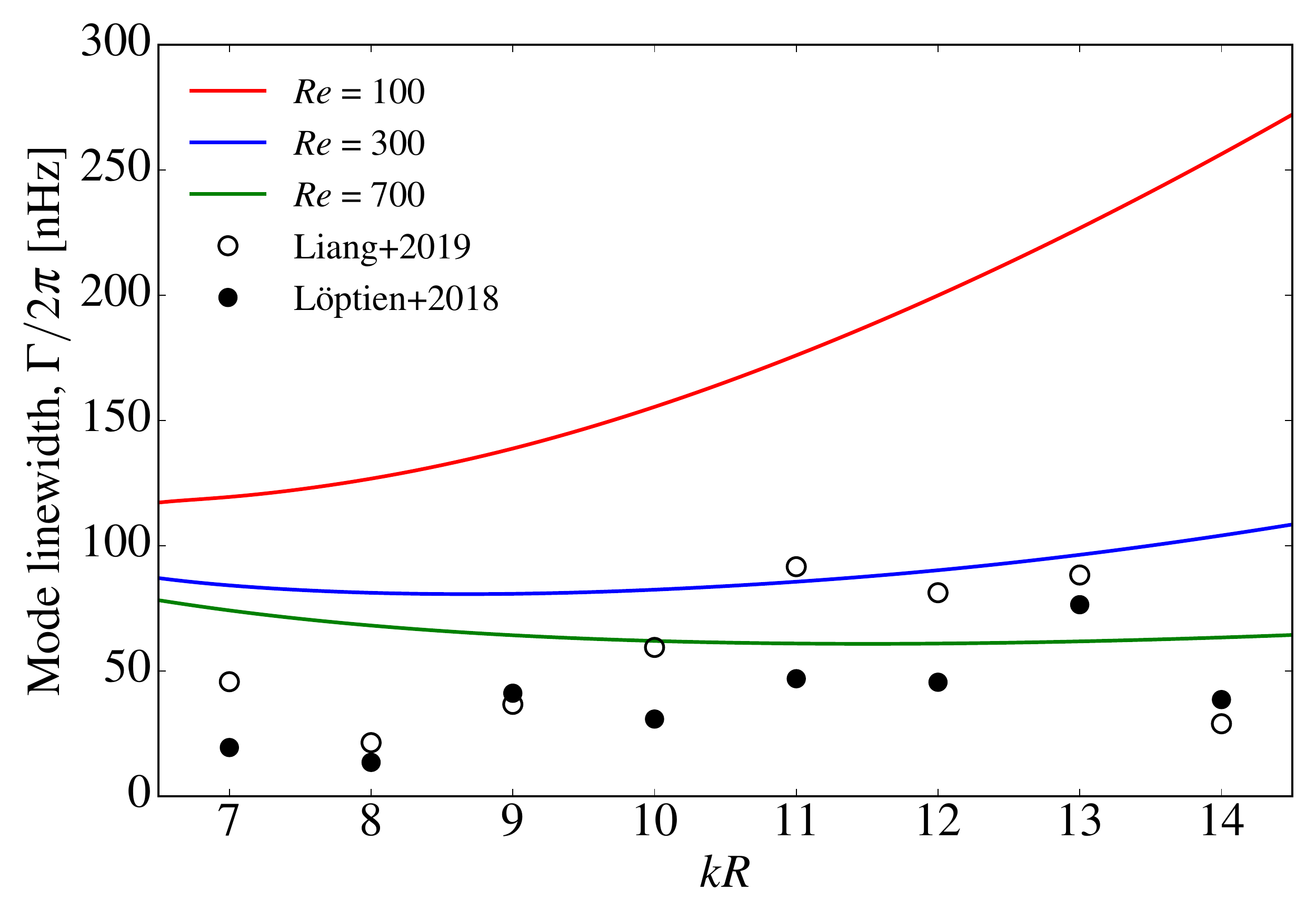}
       \caption{Left panel: R-mode dispersion relations $\omega =k c_r$ for $Re=100$, 300 and 700 (red, blue and green). The dashed black curve is the reference dispersion relation $\omega=k c_0$.
       \LG{
       For comparison, the frequencies of  each  mode $m$ observed by \citet{LOE18} and \citet{LIA19} are multiplied by the factor  $(m + 1) / kR$ and plotted at abscissa $kR$  (this simple conversion factor is derived from the  dispersion relations for classical Rossby waves in spherical and local Cartesian geometries). 
       Right panel: Plot of   $\Gamma = -2 k c_i$ for $Re=100$, 300 and 700. The observed full widths at half maximum for each mode $m$ are plotted at abscissa $kR$ for comparison.}
       }
       \label{fig:alleigenvalues}
\end{figure*}

\begin{figure*}
       \centering
       \includegraphics[width=\linewidth]{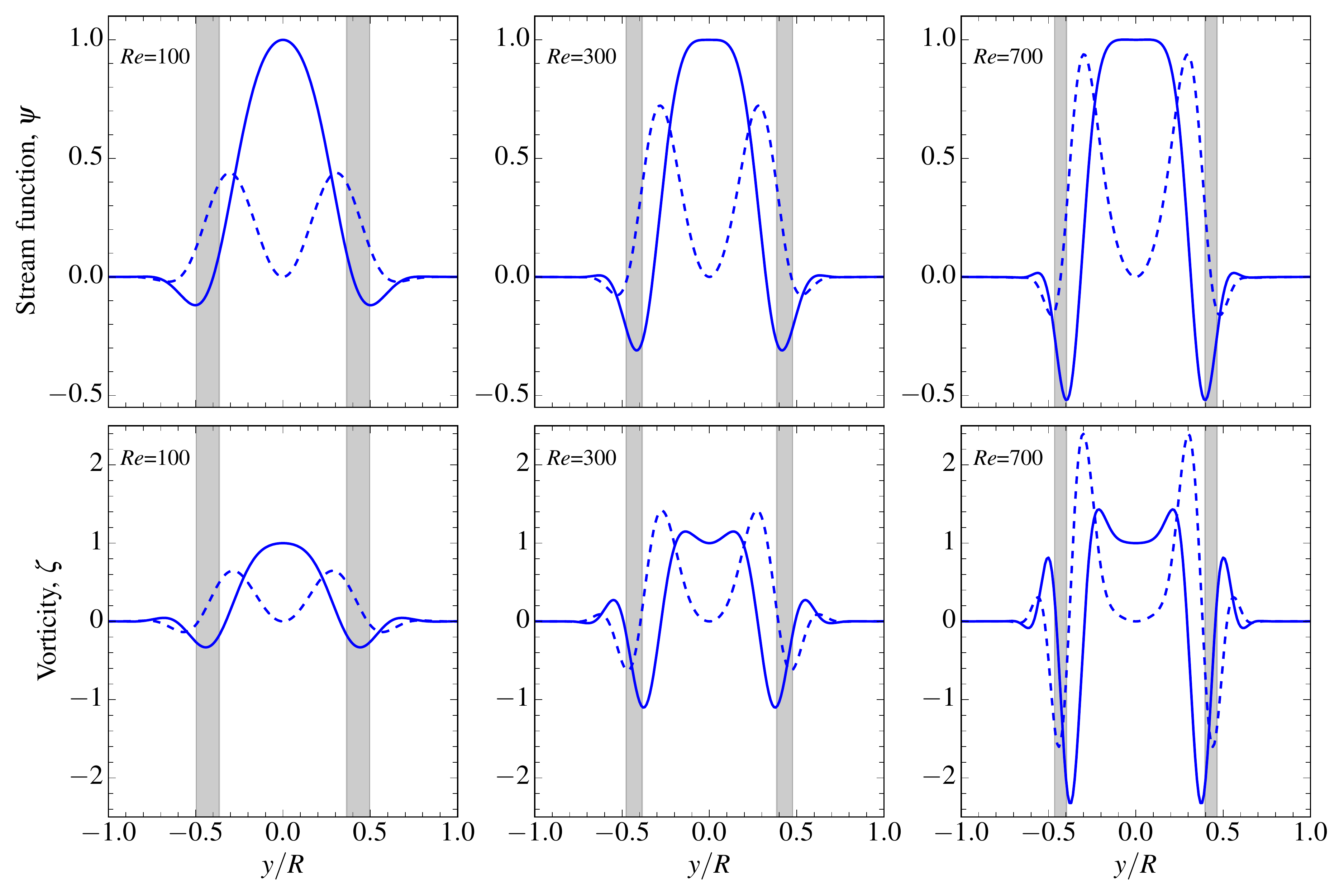}
       \caption{Stream function (top row) and vertical vorticity (bottom row) for R modes with $kR=10$. The Reynolds number is  $Re=100$, 300 and 700 from left to right. The solid and dashed curves correspond to the real and imaginary parts.
       The shaded areas indicate the locations  of the viscous critical layers for each value of $Re$.
       }
       \label{fig:R-eigenfunction}
   \end{figure*}

\subsection{Spectrum}
In Fig.~\ref{fig:eigenvalues} the eigenvalues $c=c_r + \ii c_i$ are shown in the complex plane for $\kappa = 10$ and $Re = 300$. In the figure, these eigenvalues are normalized by  $|c_0| > 0$. All modes are stable since none of the imaginary parts of the eigenfrequencies are positive. 
In the complex plane, the eigenvalues are distributed along three main branches that correspond to different types of eigenfunctions.
The same branches appear in the standard plane Poiseuille problem; they have been called  A, P and S by \citet{Mack1976}. Example eigenfunctions are displayed in Fig.~\ref{fig:eigenfunctions} and Fig.~\ref{fig:chevrons}.
The A branch, for which the eigenfunctions have large amplitudes at high latitudes, refers to the ``wall modes''.  The P branch refers to the ``center modes'' which oscillate near the viscous layers.  The  S branch corresponds to the ``damped modes'' \citep{Schensted1961}.  \citet{Schensted1961} showed that the A and P branches have a finite number of eigenvalues and the S branch has an infinite number of eigenvalues. She obtained  approximate equations for the three branches. We labelled the modes with integers in Fig.~\ref{fig:eigenvalues}, such that even integers refer to the symmetric eigenfunctions and odd integers to the antisymmetric eigenfunctions. As noted by \citet{Drazin}, the even and odd modes in the A branch have nearly the same eigenfrequencies. As seen in Fig.~\ref{fig:chevrons}, the A modes have significant amplitudes only at latitudes above the viscous layers. 

Our problem differs from the standard plane Poiseuille problem through the $\beta$ term. As a result, one additional mode appears  in the eigenvalue spectrum (Fig.~\ref{fig:eigenvalues}). This mode, which we call the R mode for an obvious reason, is symmetric and has an eigenfrequency whose real part is close to that of the classical equatorial Rossby mode, $c_r \approx c_0$.  The R mode is the mode with the longest lifetime in the spectrum. It is also the mode for which $\int_{-1}^{1} | \hat{\psi}''|^2 \id \xi$ is the smallest.
As seen in Fig.~\ref{fig:eigenfunctions}, the real part of the R-mode  eigenfunction resembles the eigenfunction of the symmetric mode found in the inviscid case (Fig.~\ref{fig:inviscidS}), except that it is smooth everywhere (no infinite derivative at the critical points). 
In the viscous case, the complex R-mode eigenfunctions look like chevrons in real space (Fig.~\ref{fig:chevrons}).

Notice that there are modes  in the P branch with $c_r$ close to $c_0$ (P$_3$ and P$_4$), however these modes have a much shorter lifetime than the R mode  (by a factor of two to three at $Re=300$) and eigenfunctions that differ significantly from the observations.

\subsection{R modes}
\label{sec:rmodes}

Figure~\ref{fig:alleigenvalues} shows the R-mode eigenfrequencies  as a function of  wavenumber $k \rsun$ for different values of the viscosity. The value of the viscosity has a rather small effect on the dispersion relation. At fixed wavenumber, the real part of the eigenfrequency changes with $Re$ by less than ten percent over the range $100 \ge Re \ge 700$. On the other hand, the imaginary part $c_i$ changes significantly with the value of $Re$. For $Re < 700$, the attenuation $\Gamma=-2 k c_i$ increases with $k$. For $Re=300$, we find that the theoretical mode linewidths  ($\Gamma / 2 \pi = -k c_i /  \pi$ in nHz) are in the range  $70$\,--\,$100$~nHz, i.e. in reasonable agreement with the observed mode linewidths \citep[$\Gamma/2\pi$ from][]{LIA19}.  Note that a mode's $e$-folding lifetime is given by \LG{$\tau=2/\Gamma$}.

The top row of Fig.~\ref{fig:R-eigenfunction}  shows the R-mode stream functions for different values of the Reynolds number. The normalization is such that, at $\xi=0$, the real part is one  and the imaginary part is zero. As $Re$ decreases, the stream function varies more slowly around the viscous layer  and its imaginary part gets smaller in amplitude. 
The bottom row of Fig.~\ref{fig:R-eigenfunction} shows the vertical vorticity eigenfunctions. Notice the fast variations near the viscous layer for $Re=700$, where  $\psi''$ is largest.

\section{Influence of the meridional flow}
\label{sec:merid}

In addition to the rotational shear $U$, here we include  the effect of the meridional flow $V$ on the R mode.  
The total velocity is
\begin{equation}
     U(y) \xhat + V(y) \yhat + \bu (x,y,t) ,
\end{equation}
where the meridional flow is approximated by
\begin{equation}
 V (y) =   \overline{V}\ \frac{y}{R}   \left[1-\left(\frac{y}{R}\right)^2 \right] ,
\end{equation}
with
\begin{equation}
\overline{V} =  \frac{3\sqrt{3}}{2} \times 15\ \text{m/s} . 
\end{equation}
The value of $\overline{V}$ is chosen such that the maximum value of $V$ is 15~m/s (near latitude $\lambda=35^\circ$).

\begin{figure}[t]
    \centering
    \includegraphics[width=\linewidth]{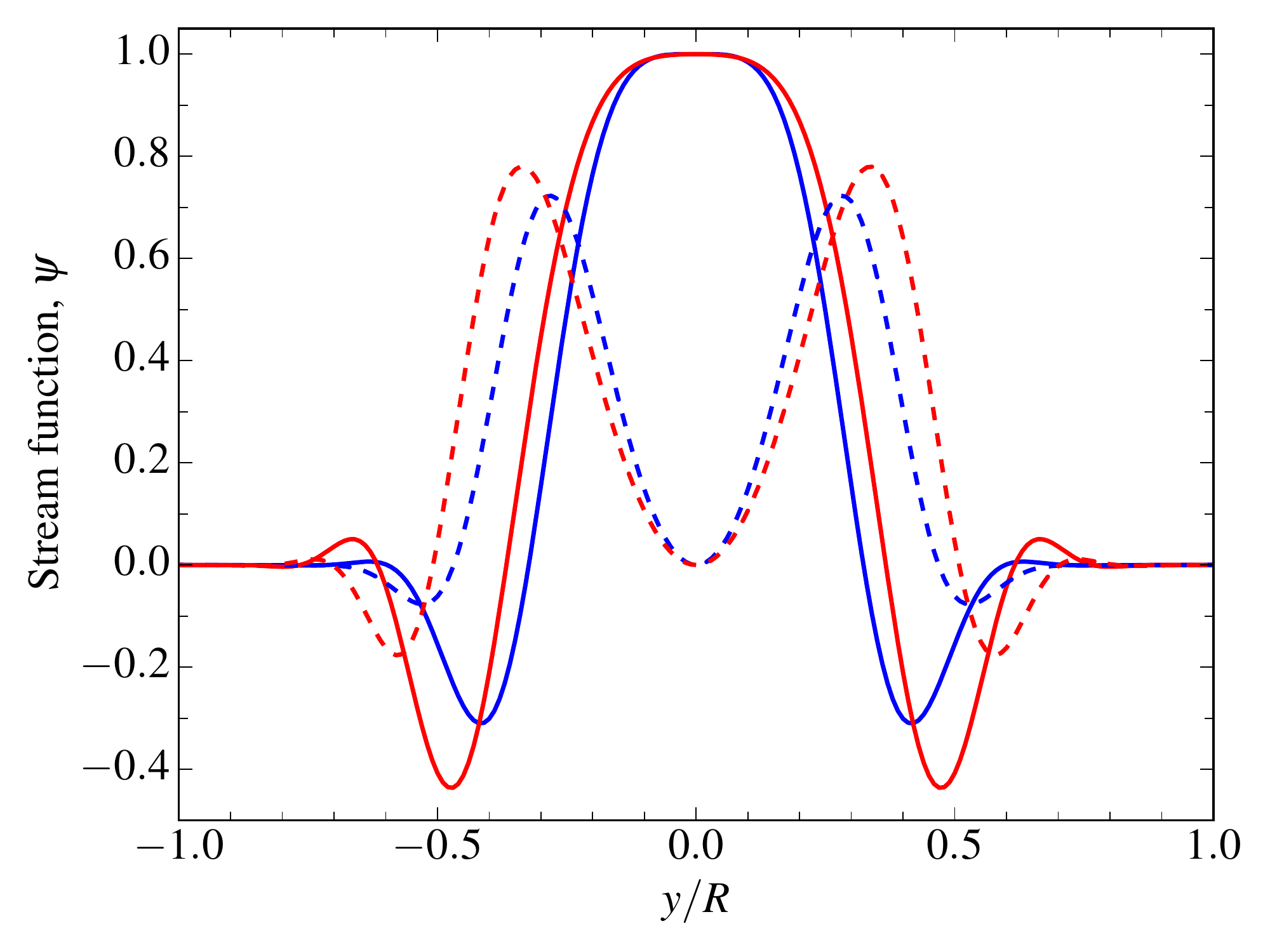}
    \caption{
    Effect of the meridional flow $V$ on the R-mode stream function for $Re = 300$ and $kR = 10$ (red curves). For comparison, the blue curves show the stream function when only the zonal flow $U$ is included. The real and imaginary parts correspond to the solid and dashed curves.
    }
    \label{fig:meridional}
\end{figure}

Including the meridional flow, the two-dimensional linearized Navier-Stokes equations in the equatorial $\beta$ plane become
\begin{align}
& \left(\frac{\partial}{\partial t}  + U \frac{\partial}{\partial x} + V \frac{\partial}{\partial y} \right)  u_x + u_y U'   = - \frac{1}{\rho} \frac{\partial p}{\partial x} + \nu \Delta u_x 
+ f u_y
,
\\
& \left( \frac{\partial}{\partial t}  +  U \frac{\partial}{\partial x} + V \frac{\partial}{\partial y} \right)  u_y  + u_y V'   = - \frac{1}{\rho}\frac{\partial p}{\partial y}  + \nu  \Delta u_y  - f u_x.
\end{align}
Combining these equations, the fourth-order differential equation for the stream function (linear problem) is
\begin{equation}
\left(c -  U +  \frac{\ii V'}{k}  \right) D \psi +   \frac{\ii V}{k} D \psi'
  -  \left( \beta - U'' \right) \psi
=    \frac{ \ii  \nu}{k} D^2  \psi,
\end{equation}
with $D=-k^2+\id^2/\id y^2$. Compared to the previous problem with $U$ only, the  term in front of $D\psi$ is now complex and an additional term involving the first and third derivatives of the stream function appear. 

We consider only the symmetric solutions and focus on the R mode.  The boundary conditions are $\psi'(0)=\psi'''(0)=0$ and $\psi(R) =\psi'(R)= 0$. Like before, we follow the procedure by \citet{ORS71} to solve the eigenvalue problem.
The eigenfrequencies are not affected significantly by the meridional flow. For $Re=300$ and $kR=10$, we find  $c/|c_0| = -0.986 -0.446  \ii$ when $U$ and $V$ are included, compared to $c/|c_0| = -0.982 -  0.398  \ii$ when only $U$ is included. Figure~\ref{fig:meridional} shows the real and imaginary parts of the R-mode eigenfunction at fixed $kR=10$ for $Re =300$. 
The meridional flow $V$ has a small but measurable influence on the $\psi$ eigenfunction: it is stretched towards higher latitudes and its real part has a larger amplitude near the viscous layers.

\begin{figure*}
       \centering
       \includegraphics[width=\linewidth]{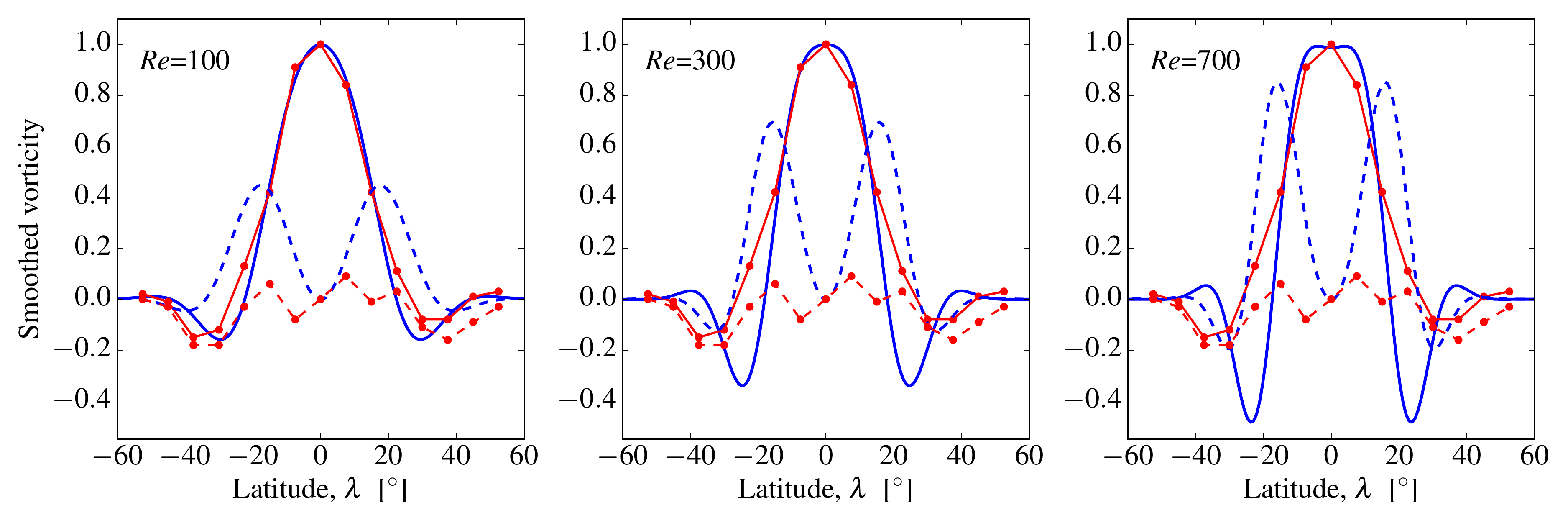}
       \caption{Real (blue solid lines) and imaginary (blue dashed lines) parts of the R-mode vertical vorticity at $kR=10$ after  smoothing the maps of horizontal velocities  with a 2D Gaussian kernel with $\sigma=6^\circ$. Three different values of $Re$ are shown. 
        For comparison, the red curves with points show the ring-diagram helioseismic observations for $m=10$ near the surface \citep{PRO20}.
       }
       \label{fig:R-smoothed-vort}
   \end{figure*}

\section{Comparison with observations}
\label{sec:obs}

\citet{PRO20} measured and characterized the vorticity eigenfunctions of the solar Rossby modes using ring-diagram helioseismic analysis. 
To enable a direct comparison between observations and theory, some smoothing must be applied to the model  because the observed flows have a resolution of only $\sigma=6^\circ$ in $\phi$ and $\lambda$.
After remapping the theoretical flows on a longitude-latitude grid, we convolve  $u_\phi$ and $u_\lambda$ with  a 2D Gaussian kernel with standard deviation $\sigma$ in both coordinates. From the smoothed velocities, we compute the vertical vorticity. As seen in Fig.~\ref{fig:R-smoothed-vort}, the smoothing has a significant effect on the theoretical vorticity near the viscous layer.

The functional form of the real part of the observed eigenfunction was described by several parameters by \citet{PRO20}: a full width at half maximum $W$, the latitude $\lambda_0$ where the sign changes, the latitude $\lambda_{\rm min}$ where the vorticity is most negative,  and the value $\zeta_{\rm min}$ of the vorticity at $\lambda_{\rm min}$.
These parameters are provided in Table~1 for three different values of $kR$ in three different cases: (a) smoothed theoretical vorticity when the zonal shear flow $U$ is included, (b) smoothed theoretical vorticity when $U$ and $V$ are both included and  (c) observations from \citet{PRO20}. For  $Re=300$,
 we find that the widths for cases (a) and (b) are within $\approx 3^\circ$ of the observed values. For (a), the zero-crossing latitude  varies from $\lambda_0=17^\circ$ for $kR=12$ to $\lambda_0=21^\circ$ for $kR=8$. For (b) the values of $\lambda_0$ are slightly higher by $\approx 2^\circ$, while the observed  $\lambda_0\approx 28^\circ$ does not vary much with $k$. The theoretical values of $\lambda_{\min}$ vary with $kR$ a little faster  than $\lambda_0$, and theory is in better agreement with observations for $kR=8$.
The observed values of $\zeta_{\rm min}$ range from  $-0.3$ to $-0.1$, about half of the theoretical values. However, the values of $\zeta_{\rm min}$ depend on $Re$, with more negative values for larger values of $Re$.
On the other hand,  the values of $W$, $\lambda_0$, and $\lambda_{\rm min}$ in the table are not very sensitive to the Reynolds number for $100\le Re \le 700$.

The imaginary part of the vorticity eigenfunctions is a lot more difficult to measure \citep{PRO20}. It is significantly different from zero for some values of $m$, however it is noisy and its functional form cannot be described by a simple parametric function. According to \citet{PRO20}, the sign of the observed imaginary part appears to be positive for the smallest values of $m$ and negative for $m >5$.  The comparison provided in Fig.~\ref{fig:R-smoothed-vort} shows that, at latitudes below the viscous layer, the amplitude of the imaginary part is much higher  in the model than in the observations.

\begin{table}
    \caption{Parameters $W$, $\lambda_0$, $\lambda_{\rm min}$ and $\zeta_{\rm min}$ characterizing the functional form of the real part of the R-mode vorticity eigenfunctions. The smoothed eigenfunctions from theory ($Re=300$) are given for the cases when (a) only the  zonal flow $U$ is included and (b) both $U$ and the meridional flow $V$ are included. The theoretical values are compared with the observations from \citet{PRO20}.}
        \centering
    \begin{tabular}{lllll}
    \hline \hline
      & Case & $kR=8$& $kR=10$ & $kR=12$  \\ \hline
      $W$ $(^\circ)$ &  $U$ & $17.6$ & $13.1$ & $11.2$\\
     &  $U$ \& $V$  & $20.2$ & $15.8$ &  $13.1$ \\
         &  Obs. & $17.1 \pm 0.8$ & $14.7 \pm 1.1$ & $13.8 \pm 1.4$\\
         \hline
      $\lambda_0$ $(^\circ)$ &  $U$  & $21.1$ & $18.4$ & $17.5$\\
     & $U$ \& $V$  & $24.8$ & $20.2$ & $19.4$ \\
         &  Obs. & $25.8$ &  $27.8 \pm 1.6$ &  $28.9 \pm 1.0$ \\
         \hline
      $\lambda_{\rm min}$ $(^\circ)$  &  $U$  & $29.3$ & $24.8$ & $22.9$\\
     &  $U$ \& $V$  & $32.9$ & $27.4$ & $24.8$ \\
         &  Obs. &  $36.5 \pm 1.8$ &  $34.5 \pm 2.1$ &  $39.8 \pm 1.8$ \\
         \hline
      $\zeta_{\rm min}$ &  $U$  & $-0.69$ & $-0.35$ & $-0.17$ \\
     &  $U$ \& $V$  & $-0.86$ & $-0.49$ & $-0.28$ \\
         &  Obs. &  $-0.28$ &  $-0.11$ &  $-0.14$ \\ 
        \hline
    \end{tabular}
\end{table}

\section{R-mode momentum fluxes}
\label{sec:reynolds}

The complex eigenfunctions in our model imply that  R modes transport  a net momentum flux in latitude. It is interesting to obtain an estimate of this momentum flux (even though the model eigenfunctions have imaginary parts that differ from  the observations). The purpose of this exercise is to establish whether R modes play a significant role in the balance of forces that shape differential rotation.

Let us consider a time-dependent zonal flow $U(y,t)$, which evolves slowly according to the $x$-component of the momentum equation averaged over $x$ \citep[see, e.g.,][]{Geisler1974}:
\begin{equation}
    \frac{\partial  U  }{\partial t}  =  
    - \frac{\partial}{\partial_y} \langle  u_x u_y  \rangle 
    + \nu  U'' + V(\beta y -U') ,
    \label{eq:Uevolution}
\end{equation}
where the prime denotes a $y$-derivative and angle brackets $\langle\cdot\rangle$ denote the average over $x$. We used $\partial_x u_x + \partial_y u_y = 0$ to obtain the term involving  $\langle u_x u_y \rangle$ on the right-hand side of the equation. 
The various terms in Eq.~(\ref{eq:Uevolution}) have been discussed in detail by \citet{Ruediger1989} in spherical geometry. These terms must balance exactly to explain the steady-state differential rotation.

\begin{figure}
       \centering
\includegraphics[width=\linewidth]{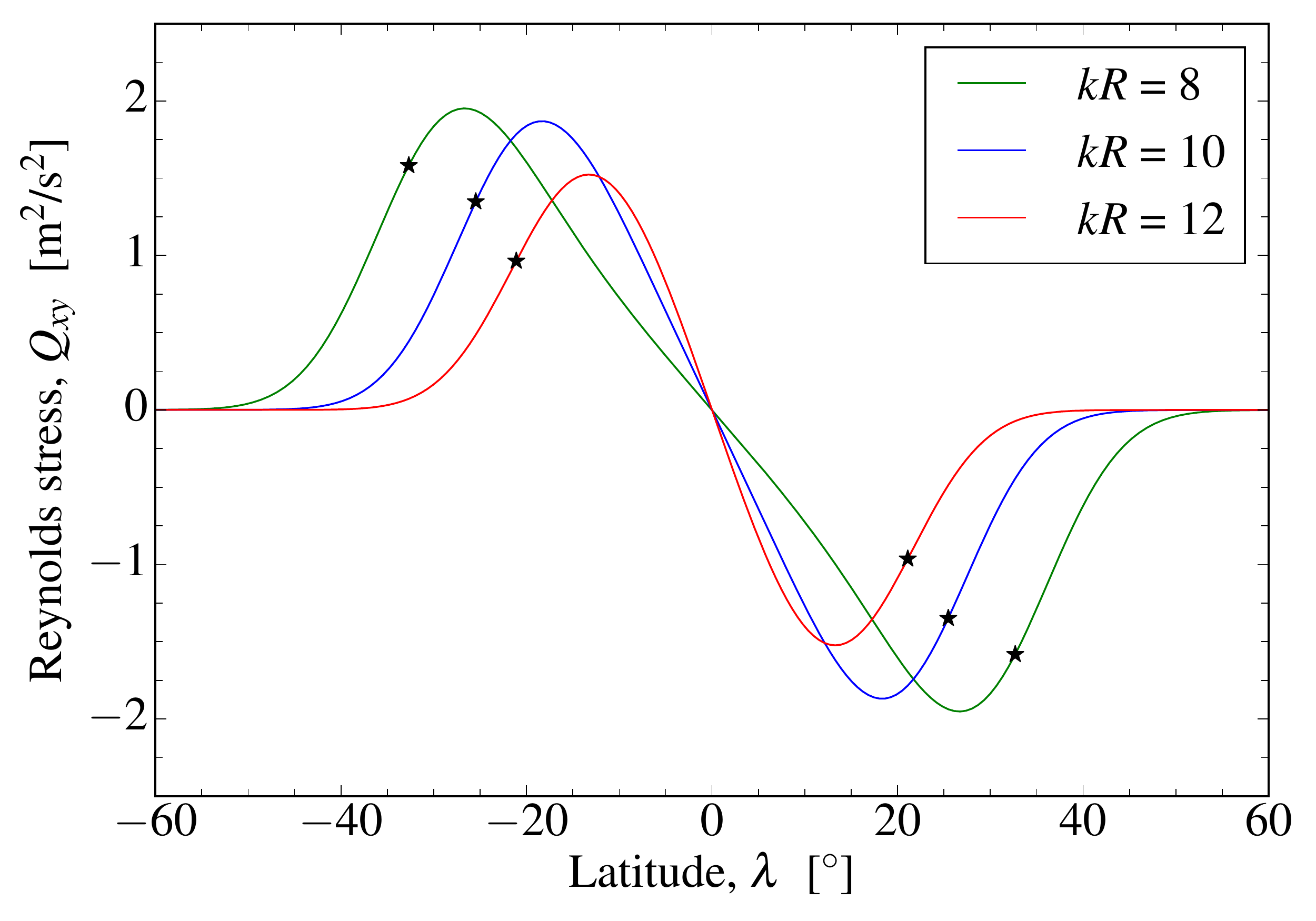} 
\caption{Horizontal Reynolds stress $Q_{xy}$ versus latitude  for R modes with $kR=8$, 10 and 12. The Reynolds number is $Re=300$ and mode amplitudes were normalized according to Eq.~(\ref{eq.normalizemode}). Both the zonal flow $U$ and the meridional flow $V$ were included to compute the mode eigenfunctions. The stars show the locations of the viscous critical layer for the different values of $kR$.}
       \label{fig:reynolds}
   \end{figure}

In our problem the horizontal Reynolds stress has two components, the first one due to rotating  turbulent convection (the $\Lambda$ effect) and the other one due to the Rossby waves,
\begin{equation}
    \langle  u_x u_y  \rangle = 
    \langle  u_x u_y  \rangle_{\rm R}
    +
    \langle  u_x u_y 
\rangle_{\Lambda} .
\end{equation}
Let us estimate $\langle  u_x u_y  \rangle_{\rm R}$ in the model for a single R mode. It is related  to the stream function through
\begin{align}
    \langle  u_x u_y  \rangle_{\rm R}
    & \LG{ = - \left\langle  \frac{\partial \Psi}{\partial y}\ \frac{\partial \Psi}{\partial x}  \right\rangle } \nonumber
    \\
      & \LG{ = - \left\langle  {\rm Re}\left(\psi'  e^{\ii k(x-c_r t) - t/\tau}\right)\
    {\rm Re}\left( \ii k \psi e^{\ii k (x-c_r t) - t/\tau}\right)   \right\rangle} \nonumber
    \\
    & =
    - \frac{k}{2} \text{ Im} \left(  \psi' \psi^*  \right) e^{- 2 t/\tau} 
    =: Q_{xy}\ e^{- 2 t/\tau} , 
\end{align}
where $\tau$ is the mode lifetime mentioned in Sect.~\ref{sec:rmodes}.
We normalize the R-mode stream function such that the velocity amplitude at the equator is equal to its observed value,
\begin{equation}
   k \psi (0) = u_{\rm max} \approx 2 \textrm{ m/s}.
   \label{eq.normalizemode}
\end{equation}
 The Reynolds stress $Q_{xy}$ is plotted in Fig.~\ref{fig:reynolds} for R modes with $Re=300$ and $kR=8$, 10 and 12.
We find that $Q_{xy} <0$ in the north, below the viscous critical layer. For example, for $kR=10$,  $Q_{xy}$ reaches the minimum value of $-2$~m$^2$/s$^2$ at latitude $20^\circ$.
This means that  R modes transport angular momentum from the dissipation layer to the equator, i.e. reinforce latitudinal differential rotation. 
\LG{This is the expected result for idealized Rossby waves incident on a critical layer \citep[see, e.g.,][]{VAL06}. }

\begin{figure}
       \centering
\includegraphics[width=\linewidth]{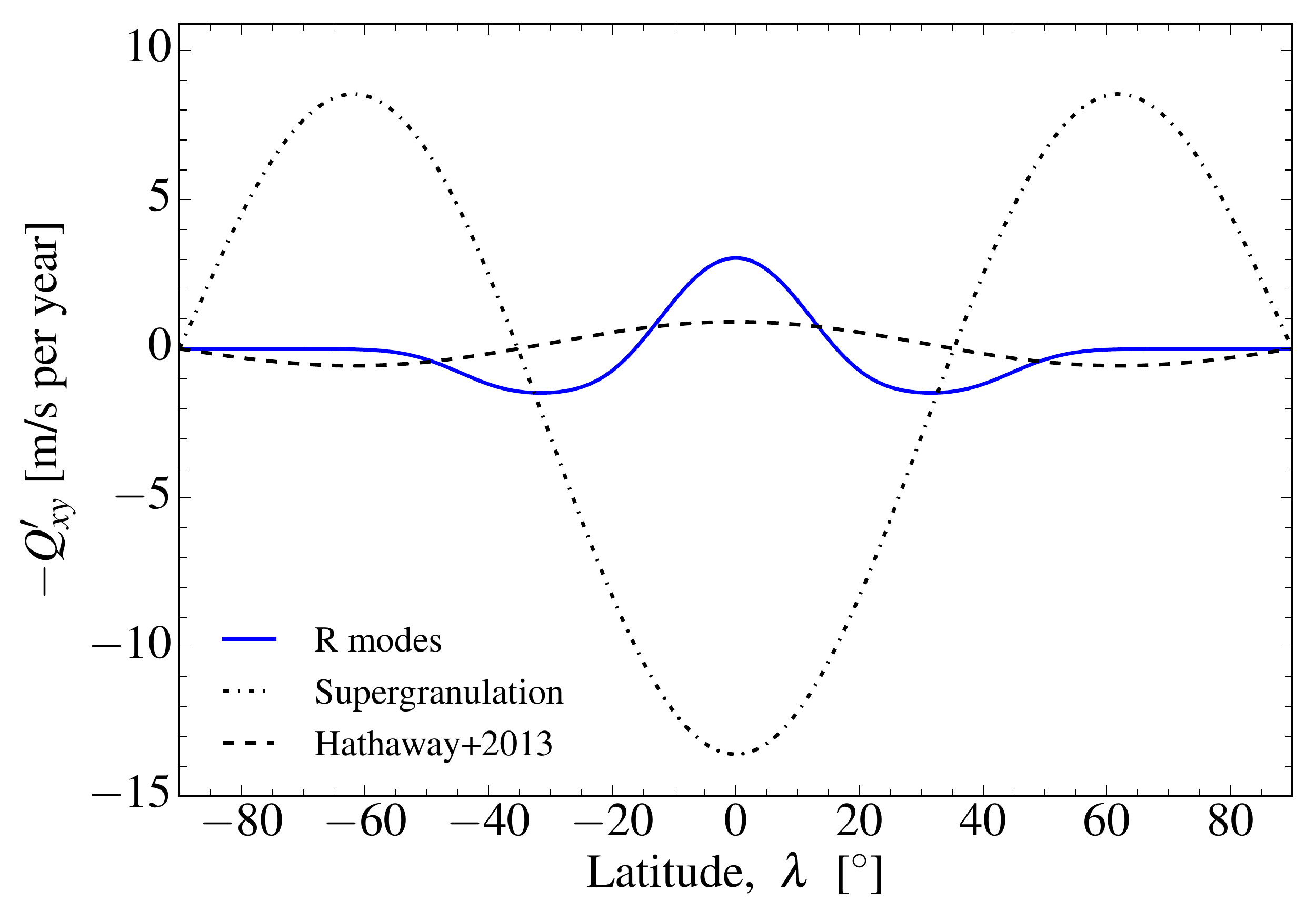} 
\caption{Equatorial acceleration $ - \partial_y Q_{xy}$ due to the superposition of nine viscous R modes for $kR=7, 8, \dots 15$ (blue solid curve). The dot-dashed and dashed curves show $-\partial_y\langle u_x u_y \rangle$ 
for supergranulation \citep[][their figure~10]{ARAA2016}  and 
larger-scale convection \citep{Hathaway2013}  respectively. For reference,  the solar ``torsional oscillation''  has a typical amplitude of about $\pm 5$ m/s
\citep[see, e.g.,][]{Lekshmi2018}.}
       \label{fig:DeltaU}
   \end{figure}

Summing $Q_{xy}$ over several R modes would lead to a horizontal Reynolds stress that is comparable in amplitude and sign to the value reported at large spatial scales by \citet{Hathaway2013}. On the other hand, $Q_{xy}$ has the opposite sign and is much smaller in amplitude than the (viscous) Reynolds stress associated with convective flows at supergranulation scales  \citep[][their figure~10]{ARAA2016}.
The acceleration $-\partial_y Q_{xy}$ is plotted in Fig.~\ref{fig:DeltaU} and compared to the above mentioned  measurements. We see that Rossby waves  in our model contribute to the  equatorial acceleration at a significant level.

\section{Conclusion}

Using a simple two-dimensional setup in the $\beta$ plane,
we have shown that latitudinal differential rotation and viscosity  must play an important role in shaping the horizontal eigenfunctions of global-scale Rossby modes. Viscous critical layers form around latitudes where $c=U$. We find that only one symmetric mode, which we called the R mode,  has an eigenvalue whose real part is close to that of the classical sectoral Rossby mode and whose imaginary part is close to the observed value when $Re\approx 300$. The real part of the vorticity eigenfunctions can be made to agree qualitatively with solar observations (unlike the imaginary part).

\LG{
Treating this problem as a stability problem for a  viscous Poiseuille flow in the $\beta$ plane enabled us to connect to prior results in the fluids literature. For example, we used a well-established method to solve the Orr-Sommerfeld equation numerically and we easily identified the known families of modes A, P  and S in the complex plane (eigenvalues and eigenfunctions). A new aspect of our work is the identification of the viscous R mode, due to the $\beta$ term in the equation.
We find that the combination of the shear flow and the viscosity lead to chevron-shaped eigenfunctions, and thus to non-zero angular momentum transport by R modes. 
In our model, horizontal Reynolds stresses due to R modes lead to significant equatorial acceleration.
Another original aspect of our work is the study of the influence of the solar meridional flow on R modes. We found that the meridional flow affects the eigenfunctions to measurable levels. 
Reynolds stresses have significantly larger amplitudes when the meridional flow is included, although the meridional flow plays a  much smaller role than the differential rotation in shaping the eigenfunctions.
}

\LG{
Sophisticated 3D models (but without background shear flow) also include Rossby waves as a possible  mechanism to produce equatorial super-rotation in the atmospheres of  planets \citep[e.g.,][]{Liu2011, Read2018} and exoplanets \citep[e.g.,][]{Showman2011}.}
Clearly, the present work will have to be extended to three dimensions \citep[see, e.g.,][for the eigenvalues in the inviscid case]{Watts2004} to account for the radial gradients of solar rotation  measured by helioseismology. Also, a better understanding of Rossby waves will benefit from more realistic numerical experiments \citep[see][]{Bekki2019}.
Finally, we note that the model developed here can be used to estimate the temporal changes in the Rossby wave frequencies due to the solar-cycle variations in the zonal flows \citep{Goddard2020}.

\begin{acknowledgements}
We thank Aaron Birch, Shravan Hanasoge (CSS), and  Bastian Proxauf for useful discussions.
{An early study of the inviscid problem may be found in the  Bachelor's thesis of Leonie Frantzen.}
Author contributions: L.G. proposed and designed research. L.G. and M.A. solved the inviscid problem analytically. D.F. solved the viscous problem numerically. L.G. wrote the draft paper. All authors reviewed the final manuscript. 
Funding: L.G. acknowledges partial support from ERC Synergy Grant WHOLE~SUN 810218 and NYUAD Institute Grant G1502.
M.A. acknowledges funding from the  Volkswagen Foundation
and the Shota Rustaveli National Science Foundation of Georgia (SRNSFG  grant N04/46).
The computational resources were provided by the German Data Center for SDO through grant 50OL1701 from the German Aerospace Center (DLR). 
\end{acknowledgements}

\bibliography{biblio}{}
\bibliographystyle{aa}

\end{document}